\documentclass[twoside,twocolumn,9pt]{article}
\usepackage{extsizes}
\usepackage[super,sort&compress,comma]{natbib}
\usepackage[version=3]{mhchem}
\usepackage[left=1.5cm, right=1.5cm, top=1.785cm, bottom=2.0cm]{geometry}
\usepackage{balance}
\usepackage{times,mathptmx}
\usepackage{sectsty}
\usepackage{graphicx}
\usepackage{lastpage}
\usepackage[format=plain,justification=justified,singlelinecheck=false,font={stretch=1.125,small,sf},labelfont=bf,labelsep=space]{caption}
\usepackage{float}
\usepackage{graphicx}
\usepackage{amsmath,amssymb}
\usepackage{bm}
\usepackage{fancyhdr}
\usepackage{fnpos}
\usepackage[english]{babel}
\usepackage{array}
\usepackage{color}
\usepackage{charter}
\usepackage[T1]{fontenc}
\usepackage[usenames,dvipsnames]{xcolor}
\usepackage{setspace}
\usepackage[compact]{titlesec}
\usepackage{epstopdf}
\definecolor{cream}{RGB}{222,217,201}
\begin{document}

\pagestyle{fancy}
\thispagestyle{plain}
\fancypagestyle{plain}{

\fancyhead[C]{\hspace{-0.5cm}\includegraphics[width=18.5cm]{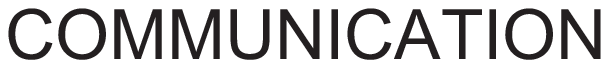}}
\fancyhead[L]{\hspace{-0.1cm}\vspace{.75cm}\includegraphics[height=50pt]{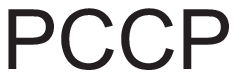}}
\renewcommand{\headrulewidth}{0pt}
}
\makeFNbottom
\makeatletter
\renewcommand\LARGE{\@setfontsize\LARGE{15pt}{17}}
\renewcommand\Large{\@setfontsize\Large{12pt}{14}}
\renewcommand\large{\@setfontsize\large{10pt}{12}}
\renewcommand\footnotesize{\@setfontsize\footnotesize{7pt}{10}}
\renewcommand\scriptsize{\@setfontsize\scriptsize{7pt}{7}}
\makeatother

\renewcommand{\thefootnote}{\fnsymbol{footnote}}
\renewcommand\footnoterule{\vspace*{1pt}%
\color{cream}\hrule width 3.5in height 0.4pt \color{black} \vspace*{5pt}}
\setcounter{secnumdepth}{5}

\makeatletter
\renewcommand\@biblabel[1]{#1}
\renewcommand\@makefntext[1]%
{\noindent\makebox[0pt][r]{\@thefnmark\,}#1}
\makeatother
\renewcommand{\figurename}{\small{Fig.}~}
\sectionfont{\sffamily\Large}
\subsectionfont{\normalsize}
\subsubsectionfont{\bf}
\setstretch{1.125} 
\setlength{\skip\footins}{0.8cm}
\setlength{\footnotesep}{0.25cm}
\setlength{\jot}{10pt}
\titlespacing*{\section}{0pt}{4pt}{4pt}
\titlespacing*{\subsection}{0pt}{15pt}{1pt}

\fancyfoot{}
\fancyfoot[RO]{\footnotesize{\sffamily{1--\pageref{LastPage} ~\textbar  \hspace{2pt}\thepage}}}
\fancyfoot[LE]{\footnotesize{\sffamily{\thepage~\textbar\hspace{0.cm} 1--\pageref{LastPage}}}}
\fancyhead{}
\renewcommand{\headrulewidth}{0pt}
\renewcommand{\footrulewidth}{0pt}
\setlength{\arrayrulewidth}{1pt}
\setlength{\columnsep}{6.5mm}
\setlength\bibsep{1pt}

\makeatletter
\newlength{\figrulesep}
\setlength{\figrulesep}{0.5\textfloatsep}

\newcommand{\topfigrule}{\vspace*{-1pt}%
\noindent{\color{cream}\rule[-\figrulesep]{\columnwidth}{1.5pt}} }

\newcommand{\botfigrule}{\vspace*{-2pt}%
\noindent{\color{cream}\rule[\figrulesep]{\columnwidth}{1.5pt}} }

\newcommand{\dblfigrule}{\vspace*{-1pt}%
\noindent{\color{cream}\rule[-\figrulesep]{\textwidth}{1.5pt}} }

\makeatother

\twocolumn[
  \begin{@twocolumnfalse}
\vspace{0cm}
\sffamily
\begin{tabular}{m{4.5cm} p{13.5cm} }

{\vspace{70.pt}\includegraphics[height=92pt]{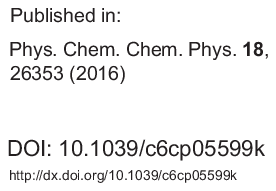}} & \noindent\LARGE{\textbf{Bidirectional particle transport and size selective sorting of Brownian particles in a flashing spatially periodic energy landscape}} \\
 & \vspace{0.3cm} \\

 & \vspace{-1.3cm}\noindent\large{ Fernando Martinez-Pedrero,\textit{$^{a}$} Helena Massana-Cid,\textit{$^{a}$} Till Ziegler,\textit{$^{b}$} Tom H. Johansen,\textit{$^{c,d}$} Arthur V. Straube,\textit{$^{a,e}$} and Pietro Tierno$^{\ast}$\textit{$^{a,f,g}$}} \\

 & \\

\end{tabular}

 \end{@twocolumnfalse} \vspace{0.6cm}

  ]

\renewcommand*\rmdefault{bch}\normalfont\upshape
\rmfamily
\section*{}
\vspace{-1cm}


\footnotetext{\textit{$^{a}$~Departament de F\'isica de la Mat\`eria Condensada, Universitat de Barcelona, Av. Diagonal 647, 08028, Barcelona, Spain. Tel: +34 9340 934034031; E-mail: ptierno@ub.edu}}
\footnotetext{\textit{$^{b}$~Department of Physics, Humboldt-Universit\"at zu Berlin, Newtonstr. 15, 12489 Berlin, Germany. }}
\footnotetext{\textit{$^{c}$~Department of Physics, The University of Oslo, P.O. Box 1048 Blindern, 0316 Oslo, Norway.}}
\footnotetext{\textit{$^{d}$~Institute for Superconducting and Electronic Materials, University of Wollongong, Wollongong, New South Wales 2522, Australia}}
\footnotetext{\textit{$^{e}$~Department of Mathematics and Computer Science, Freie Universit\"at Berlin, Arnimallee 6, 14195, Berlin, Germany}}
\footnotetext{\textit{$^{f}$~Institut de Nanoci$\grave{e}$ncia i Nanotecnologia, IN$^2$UB, Universitat de Barcelona, Av. Diagonal 647, 08028, Barcelona, Spain.}}
\footnotetext{\textit{$^{g}$~Universitat de Barcelona Institute of Complex Systems (UBICS), Universitat de Barcelona, Barcelona, Spain}}


\sffamily{\textbf{
We demonstrate a size sensitive experimental scheme which
enables bidirectional transport
and fractionation of
paramagnetic colloids in a fluid medium.
It is shown that two types of magnetic colloidal particles
with different sizes
can be simultaneously transported in opposite directions,
when deposited above a stripe-patterned
ferrite garnet film subjected to
a square-wave magnetic modulation.
Due to their different sizes, the particles
are located at distinct elevations above the surface,
and they experience two different energy landscapes,
generated by the modulated magnetic substrate.
By combining theoretical arguments and numerical simulations,
we reveal such energy landscapes, which fully explain the bidirectional transport
mechanism.
The proposed
technique does not require pre-imposed channel geometries
such as in conventional microfluidics
or lab-on-a-chip systems,
and permits remote control over the particle motion, speed and
trajectory, by using relatively low
intense magnetic fields.}}\\
\rmfamily 

The nonequilibrium dynamics
of particles driven above
periodic potentials are common to different physical
and biological systems,
from vortices in superconductors,~\cite{Coh97}
to charge density waves,~\cite{Gru88}
frictional surfaces,~\cite{Boh12}
cell migration~\cite{Mah09}
and molecular motors.~\cite{Jul97}
On the technological side,
periodic potentials generated by
patterned substrates have been successfully used
in the past to
transport and separate microscopic
particles dispersed in a fluid medium.
It is possible to direct different
particle species along
separate paths
under similar experimental conditions,
when the interaction of the
particles with the imposed landscape
depends on their particular properties,
such as their size, electric, magnetic susceptibility
or surface charge.
Particle sorting has been demonstrated in the past via electric~\cite{Reg07,Kim08,Bog12},
magnetic~\cite{Yel05,Gun05,Rob11,Lim14} and optic~\cite{Xia10,Mac13,Brz13}
fields among other types of driving mechanisms.~\cite{Xia06,Kei08}
In most of the cases,
the colloidal species are deflected
from the periodic potential,
and the degree of deflection is used to transport
different particles towards different locations.
A smaller number of experimental realizations have
demonstrated the possibility to
use an external field to direct the particles in opposite directions.
In this context, most of the examples
have been obtained by
driving the particles with an electric field,
in presence of a periodic
pattern,~\cite{Reg07,Eic10,Bog12} or using binary mixtures of oppositely charged particles.~\cite{Vis11,Vis112}
This task is even more challenging if the particles are driven with
a magnetic field. In this case the periodic potential
is usually generated by a structured magnetic substrate,
and unless a proper
strategy to steer polarizable particles is designed,\cite{Gun05,Lim14}
the magnetic particles typically exhibit
a unidirectional transport.

In this article we report on the
controlled transport and separation of paramagnetic colloidal
particles driven by an externally modulated spatially periodic
energy landscape, generated across the surface of
a structured magnetic substrate.
\begin{figure*}[ht]
\centering
\includegraphics[width=0.8\textwidth,keepaspectratio]{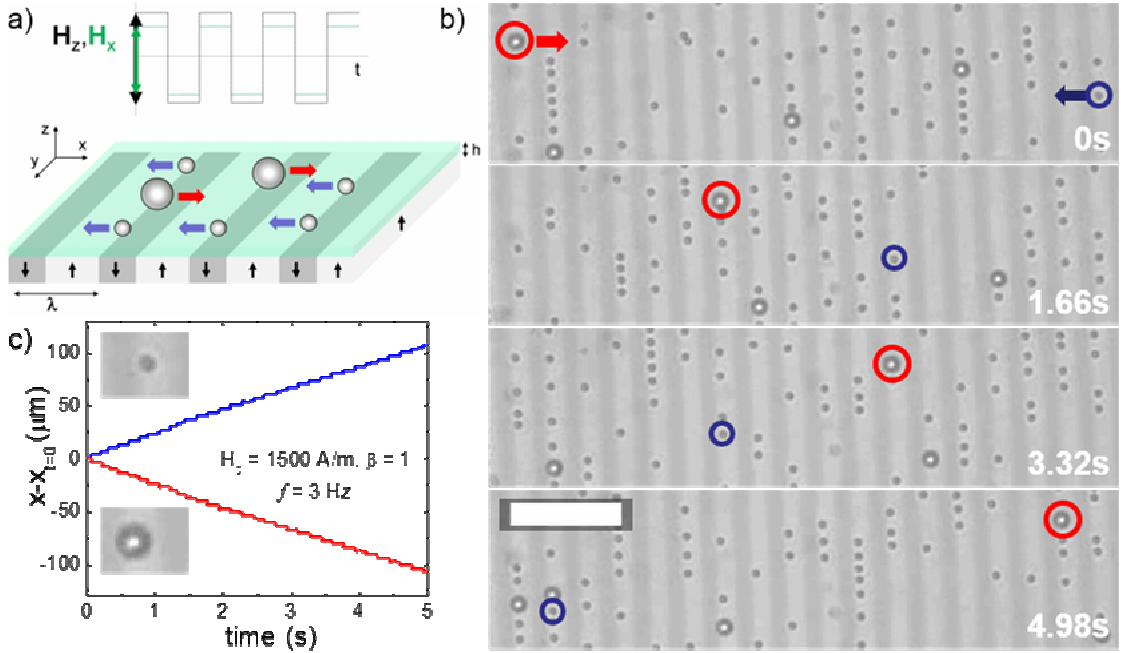}
\caption{(a) Schematic showing a binary mixture
of paramagnetic colloids deposited above
a ferrite garnet film, characterized by
parallel stripes with spatial periodicity $\lambda=6.8 \,{\rm \mu m}$.
The two applied field modulations are characterized by
strengths $H_x$ and $H_z$ of zero phase lag.
(b) Sequence of microscope images showing the bidirectional transport
of a large particle (moving to the right, size $2.8 \,{\rm \mu m}$, red circle)
and a small particle (moving to the left, size $1.0 \,{\rm \mu m}$, blue circle).
The applied field has amplitude $H_0=1500 \,{\rm A/ m}$, anisotropy $\beta=1$
and frequency $f=3 \,{\rm Hz}$.
The corresponding video-clip can be found in the Supporting Information (VideoS1).
(c) Displacements along the $x$-axis versus time of
the large (red) and small (blue) particle.}
\label{fig_1}
\end{figure*}
We demonstrate via real-time experiments
that colloidal particles
of different sizes
can be transported in opposite directions
at a well defined speed, the latter being mainly controlled by the driving
frequency of the applied field.
The mechanism behind the bidirectional transport
is rather robust,
and it is fully determined by
the different shapes
that the magnetic potential adopts at the different
particle elevations.
We use theoretical arguments
to assess the shape of these
energy landscapes,
finding a good agreement
between the experimental data
and the numerical simulations
performed using these potentials.
Extending our procedure to a polydisperse colloidal system,
we determine the sensitivity to sorting
colloidal species when transported across the magnetic substrate.

Fig.~\ref{fig_1}(a) shows a sketch of the experimental system.
The substrate potential
is generated by the stray field
of a bismuth substituted ferrite garnet film (FGF),
of composition Y$_{2.5}$Bi$_{0.5}$Fe$_{5-q}$Ga$_{q}$O$_{12}$
($q=0.5-1$).
The FGF
is grown by dipping liquid phase epithaxy,
and is characterized by parallel ferromagnetic domains
with alternating up and down magnetization and
a spatial periodicity of $\lambda=6.8\,{\rm \mu m}$.~\cite{Tie09}
Bridging the magnetic domains
are Bloch walls (BWs), i.e. $\sim 20\; {\rm nm}$ narrow
transition regions where the magnetization vector rotates
by $180$ degrees.
The position of the BWs can be easily manipulated with the
aid of a relatively low (few mT) external field,
a feature which is used to modify the
magnetic stray field of the film.
We demonstrate the
bidirectional transport
by dispersing small (s) and large (L) paramagnetic colloids
above the FGF, with diameters $d_{\rm s}=1.0 \,{\rm \mu m}$ (Dynabeads Myone, Dynal) and
$d_{\rm L}=2.8 \,{\rm \mu m}$
(Dynabeads M-270, Dynal),
and magnetic volume susceptibilities
$\chi_{\rm s}=1$ and $\chi_{\rm L}=0.4$, respectively.~\cite{Cli07,Hel07}
Both types of particles are composed of a crosslinked
polystyrene matrix with surface COOH groups and
are diluted in highly deionized water (MilliQ)
before being deposited above the FGF surface.

Once above the FGF,
the particles sediment due to density mismatch,
and are attracted by the
stray field $\mathbf{H}^{\rm sub}$ of the film,
which confines them to the horizontal plane.
To avoid particle sticking and reduce the strong attraction of the FGF,
we coat the latter with a uniform layer (thickness $h$) of
a positive photoresist AZ-1512 (Microchem, Newton,
MA) via standard spin coating and UV photo-crosslink.~\cite{Tie12}
Due to weak residual electrostatic repulsive interaction, particles hover slightly above the coating layer
without mechanically touching it.
Under no applied field,
the FGF generates a one-dimensional $\lambda/2$-periodic
magnetic landscape. The particles are attracted to the minima of the landscape
located above the BWs
and stay at fixed elevations, $z_{\rm s}$ for the small or
$z_{\rm L}$ for the large particles.
The external fields are applied with a custom-made
triaxial coil system connected to
a wave generator (TTi-TGA1244, TTi) feeding a power amplifier
(IMG STA-800, Stage Line).
The particle dynamics are visualized with an upright optical
microscope (Eclipse Ni,
Nikon) equipped with a CCD
camera (Scout scA640-74f, Basler)
operating at $60$ frames per second.

As shown in Fig.~\ref{fig_1}(a), we transport the magnetic colloids
above the FGF film upon application of a
square-wave magnetic field,
$\mathbf{H}^{\rm ac} \equiv [H_x {\rm sgn}(\cos{(2 \pi f t)}),0, H_z {\rm sgn}(\cos{(2 \pi f t)})]$,
where ${\rm sgn}(x)$ is the signum function
and $f$ denotes the driving frequency.
The field strength
can be described in terms of
the amplitude $H_0=\sqrt{(H_x^2+H_z^2)/2}$
and the field anisotropy $\beta=H_x/H_z$.
The applied field modulates the magnetic domains such that
the potential is switched between two states every half ($T/2$) a period, $T=1/f$,
inducing a net particle current above the FGF.
Before discussing in detail
the resulting potential,
we will describe the
main experimental observations.
As shown in Fig.~\ref{fig_1}(b),
we find that for a certain range of field amplitudes
and anisotropies,
particles of different size are
transported in opposite directions
above the FGF surface.
In particular,
the series of polarization microscopy images
in Fig.~\ref{fig_1}(b)
show one large particle (highlighted by the red circle) and
a small particle (highlighted by the blue circle),
which completely exchange their positions after $4.98\,{\rm s}$.
Both particles display step-like trajectories reflecting the discontinuous nature
of the driving field, see Fig.~\ref{fig_1}(c) and Supporting Information
(VideoS1).
The particle motion proceeds through a series of
successive steps: every half a period $(T/2)$,
when the oscillating field changes sign,
the particles start to move towards the nearest energy minimum (steep part of the trajectory) at a distance of
one magnetic domain ($\lambda/2$), reach it and wait (flat part of the trajectory) for a next change of sign.
This movement leads
to an average speed $\langle v_{j} \rangle= \pm \lambda f$
with $j=\{{\rm s,L}\}$.
In a more concentrated system,
particles propelling in opposite directions
might accidentally collide.
In this case,
the small particles
usually slide over the surface of the large ones and
continue further almost unperturbed.
Thus, for the discontinuous
magnetic modulation used here, the
magnetic dipolar interactions between particles of different sizes are
irrelevant, as recently observed for a
binary mixture of particles driven unidirectionally
by a continuous modulation.\cite{Tie16}
In contrast, the pair interactions between
similar particles are non-negligible, as discussed
in a previous work.\cite{Str14} However, for the measurements
of average speed, we avoid chain formation
by using low concentrations
of particles, such that the pair interactions do not affect the
particle motion.
We also note that the bidirectional transport
observed for the square-wave modulation
was not found for other
types of magnetic driving,
such as linear oscillating~\cite{Tie08} or
rotating magnetic fields.~\cite{Tie122}
\begin{figure}[t]
\centering
\includegraphics[width=\columnwidth,keepaspectratio]{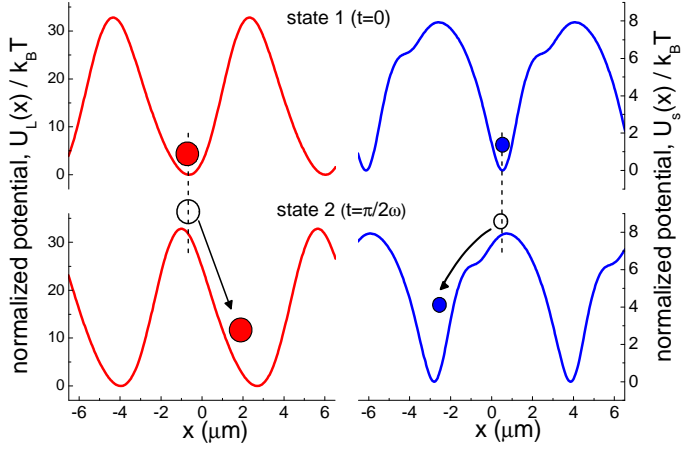}
\caption{(color online)
Normalized magnetic potential
$U_j(x)/k_BT$, $j=\{{\rm s,L}\}$
illustrating the mechanism of bidirectional transport of the
large ($2.8 \,{\rm \mu m}$, red left panels)
and the small ($1.0 \,{\rm \mu m}$, blue right panels) particles, as described in the text.
Top panels show two states of the flashing potential plotted at $t=0$ and bottom panels after
half a period at $t=T/2=\pi/2\omega$.
}
\label{fig_2}
\end{figure}

In order to understand the mechanism behind the bidirectional transport,
we calculate the magnetic energy landscape
generated by the FGF surface under the influence of an external field.
We consider two paramagnetic colloidal particles having different diameters $d_j$,
volumes $V_j=\pi d_j^3/6$, and effective magnetic volume susceptibilities $\chi_j$.
Under a magnetic field ${\mathbf H}$ the particles acquire a dipole
moment ${\mathbf m}_j=V_j \chi_j {\mathbf H}(x,z_j,t)$,
and an energy $U_j(x,t)=\frac{1}{2}(\mu_0 V_j \chi_j) \mathbf{H}^2(x,z_j,t)$.
Here, $\mu_{\rm 0}=4\pi\times 10^{-7}\,{\rm H/m}$ is the magnetic
permeability of the free space,
and the global magnetic field $\mathbf{H}=\mathbf{H}^{\rm ac}+\mathbf{H}^{\rm sub}$ is the sum of
the external modulation $\mathbf{H}^{\rm ac}$ and the substrate
field $\mathbf{H}^{\rm sub}$ expressed as~\cite{Str13}
\begin{equation}
\mathcal{H}^{\rm sub}=-\frac{2 M_s}{\pi}\log{ \left( \frac{1-u_{-}}{1-u_{+}} \right)} \,  \label{Hsub}
\end{equation}
with $H_x^{\rm sub}={\rm Re} \{ \mathcal{H}^{\rm sub} \}$
and $H_z^{\rm sub}=-{\rm Im} \{ \mathcal{H}^{\rm sub} \}$. Here, $u_{\pm}={\rm e}^{-k z}{\rm e}^{ik (x \pm \Delta)}$
with $k=\frac{2\pi}{\lambda}$ the wave number, $M_s$ is the saturation magnetization,
and $\Delta(t)=\epsilon H^{\rm ac}_z(t)/2$ is the displacement of BWs via the external field,
where $\epsilon$ is a prefactor which determines the amplitude of such displacement.
Fig.~\ref{fig_2} shows the magnetic energy landscape for the two particles
calculated at two different time steps, separated by half a period
of the driving field.
The modulation switches the potential between two states,
which are represented by the $\lambda$-periodic
profiles in the top (state 1) and bottom (state 2)
panels of the image.
Because of exponentially strong dependence on the elevation in eqn~(\ref{Hsub}),
small differences in the particle size significantly affect the shape of
the potential seen by the two types of particles.
As shown in Fig.~\ref{fig_2}, the potential for a small particle has a more complex shape
as compared to the magnetic landscape of the large one.
This shape results from higher order harmonic terms in the
Fourier expansion of eqn~(\ref{Hsub}) which decrease with the elevation, thus remaining non-negligible for the small particle
but becoming less relevant for the large one.
In both cases, a particle in state $1$ appears close to a maximum
when the switching to state $2$ occurs and tends towards the nearest minimum.
The crucial point is that the emerging maximum is always
slightly displaced with respect to the minimum,
to the left for the large and to the right
for the small particles,
forcing the two species to move in opposite directions.\cite{Cha94}
At low frequencies (such that a particle has enough time
to reach its target minimum within the time $T/2$),
its average speed $\langle v_{j} \rangle= \pm \lambda f$, as discussed above.
In contrast, when the switching between the two states is too fast,
the particle is unable to reach the minimum in such brief period of time and gets trapped,
moving back and forth between the two minima in states $1$ and $2$.
As a consequence, at high frequencies the average speed
vanishes for both types of particles.

\begin{figure}[t]
\centering
\includegraphics[width=\columnwidth,keepaspectratio]{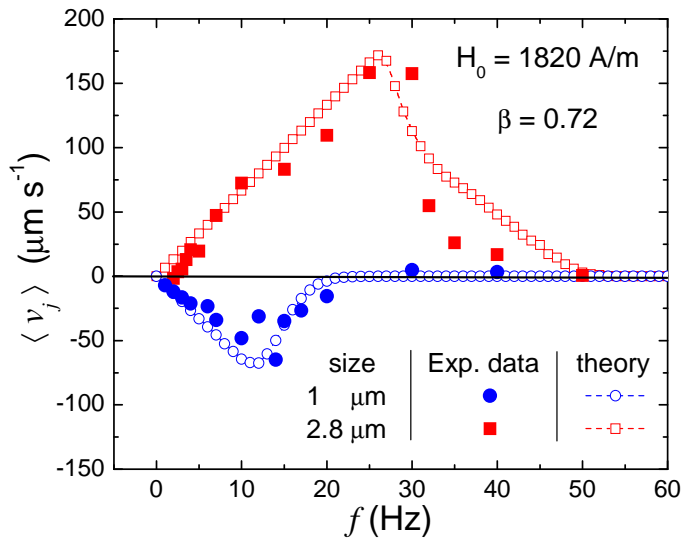}
\caption{(color online) Average speed $\langle v_j \rangle$
versus field frequency $f$ for a large (red) and small (blue)
colloidal particle driven above the FGF. The
magnetic modulation is characterized by $H_0=1820 \, {\rm A/ m}$,
and $\beta=0.72$.
Filled points denote experimental data, while
open symbols connected by dashed lines
are from the numerical simulations.}
\label{fig_3}
\end{figure}

In order to probe these theoretical arguments,
we next perform a series of experiments and
measure the average speed $\langle v_j \rangle$
of the two types of driven particles,
varying the frequency and keeping constant the amplitude
and the anisotropy of the applied field.
The experimental data in Fig.~\ref{fig_3} (the filled symbols)
show the speeds having different signs for the two types of particles.
In both cases, the
motion is characterized by two well defined regimes, as described above.
At low frequencies, the particles display net motion with the average speed
$\langle v_j \rangle$ that increases almost linearly with $f$.
In contrast, beyond a critical value ($f^{\rm c}_{\rm s}=14.2 \, {\rm Hz}$
for the small particles and $f^{\rm c}_{\rm L}=30 \, {\rm Hz}$ for the large ones),
the beads cannot follow the flashing minimum, showing a
gradual decrease of $\langle v_j \rangle$ with $f$.
To achieve a quantitative agreement with the experiment, we simulate
the individual dynamics of particles in the magnetic potential based on
eqn~(\ref{Hsub}) with account of thermal fluctuations, as described by the equation
\begin{equation}
\zeta_j \frac{d x_j}{dt} = -\frac{\partial U_j(x_j,t)}{\partial x} + \sqrt{2 \zeta_j k_{\rm B} T } \,\mathbf{\xi}_j(t)\,, \quad j=\left\{ {\rm s, L}\right\}. \label{LEs} 
\end{equation}
Here, $\zeta_j=3\pi \eta d_j$ is the viscous friction
coefficient, $\eta=10^{-3} \,{\rm Pa \; s}$ is the dynamic
viscosity of the solvent, $k_{\rm B}T$ is the thermal energy, and $\xi_j(t)$ is
a Gaussian white noise with zero mean and unit covariance.
The results of the numerical simulations, shown as empty symbols in Fig.~\ref{fig_3},
agree well with the experimental data
for all the range of frequencies
explored.
In the numerical simulation
we fixed the
elevation of the particles as $z_{\rm s} = 0.9 \;{\rm \mu m}$ ($z_{\rm L}=  1.8 \;{\rm \mu m}$)
for the small (large) particle,
considering that the thickness of the polymer film
covering the FGF surface is $h= 0.4 \;{\rm \mu m}$.
We used the corresponding experimental
values for the field amplitude $H_0=1820 \;{\rm A/m}$ and the anisotropy $\beta=0.72$.
The saturation magnetization
was a free parameter
and we find the best agreement with $M_s=27 \;{\rm kA /m}$.

\begin{figure}[t]
\centering
\includegraphics[width=\columnwidth,keepaspectratio]{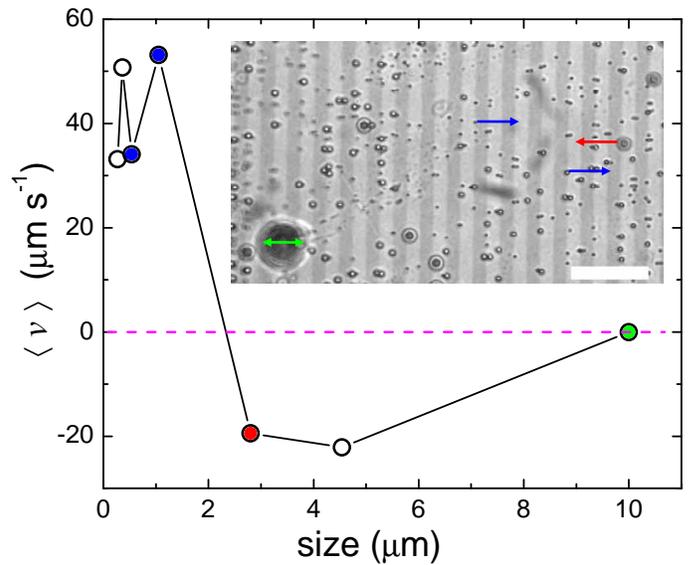}
\caption{(a)  Average speed $\langle v \rangle$ versus size
of different paramagnetic colloids
driven above the FGF
upon application of a
square-wave magnetic field
with amplitude $H_0=4300 \,{\rm A/m}$,
anisotropy $\beta=1.12$ and driving frequency
$f=12\, {\rm Hz}$.
Inset shows a snapshot
of the FGF with all types of particles,
where only three are highlighted
by arrows. In particular, the arrows
indicate the direction of motion of
particles having sizes
$1.0 {\mu m}$ (blue arrow, to the right),
$2.8 {\mu m}$ (red arrow, to the left)
and $10 {\mu m}$ (green double arrow, no average motion).
Scale bar is $20 \, {\rm \mu m}$. The corresponding video-clip (VideoS2)
can be found in the Supporting Information.}
\label{fig_4}
\end{figure}

We now explore the sorting capability of our system
by changing the size of the employed paramagnetic
particles. We deposit on the FGF a polydisperse
suspension composed
of particles having $7$ different sizes, from $270 \; {\rm nm}$
to $10 \;{\rm \mu m}$. The corresponding average speeds for a suspension subjected to a field
with amplitude $H_0=4300 \;{\rm A/ m}$, frequency $f=12 \;{\rm Hz}$
and anisotropy $\beta=1.12$, are shown in Fig.~\ref{fig_4}.
Particles having size $d\leq 1 \;{\rm \mu m}$
display a positive velocity, while
larger particles are transported in opposite direction.
We note that the largest particle ($d=10 \;{\rm \mu m}$) is unable to move on average.
For such long distance to the particle center, the landscape gets perfectly symmetric, and
the potential switches between two inverted states,
which forbids any net motion.\cite{Rei08}
Moreover, because of stronger thermal fluctuations, smaller nano-particles
exhibit a lower speed as compared to the
paramagnetic colloids having $d=1 \;{\rm \mu m}$
size. Fig.~\ref{fig_4} makes it evident that the threshold for the magnetic separation
depends crucially
on the distance between the particle center and the FGF, assuming
that each particle is doped by the same fraction of
magnetic content. One thus can increase
or reduce the thickness of the polymer film deposited above the
FGF in order to tune this threshold, allowing to efficiently sort
magnetic colloids with different sizes.

To summarize, we have experimentally demonstrated
and theoretically explained
a novel method to control and transport paramagnetic colloids
in opposite directions.
We note that
an earlier study using a different
FGF film showed an apparently similar
size-based particle separation method.\cite{Tie08}
In that work,
bidirectional motion
of a bidisperse mixture of particles
was induced
by applying a rather complicated Lissajous-like magnetic modulation,
characterized by two different driving frequencies.
In contrast, here we straightforwardly use
one driving frequency
and a switching field which
simultaneously generates
different energy landscapes depending on the elevation.
The transport method
proves robust over a wide range of
field parameters, allowing one to reach
an unprecedented
difference in the speed
between particles of different sizes,
as high as $|\langle v^L \rangle - \langle v^s \rangle| = 80 \,{\rm \mu m/ s}$.
As for fundamental implications of our work,
the realization of a flashing potential able
to simultaneously generate
opposite particle fluxes
may prompt further extensions of
previously developed ratchet models~\cite{Rei08,Han09}. In this context,
an intriguing aspect would be the role of the thermal fluctuations
in the process,
which are negligible for the large particles used in the present study,
but become increasingly important while reducing the size of the particles down to the nanoscale.
On the technological side,
our work could provide a method to fractionate
magnetic micro-spheres in a channel-free
microfluidic environment,
since it does not require
the presence of hard walls to
confine the motion of micro-particles.
Furthermore, paramagnetic particles can be easily functionalized
in order to target and bind specific
biological or chemical
agents, which could be then transported on command by the applied field.
This feature opens up the possibility for novel applications in
micro- and nanofluidic systems.

We thank L. Schimansky-Geier and I. M. Sokolov
for valuable discussions.
F. M. P., H. M. C. and P. T. acknowledge support from the ERC Starting
Grant ``DynaMO'' (Proj. No. 335040).
P. T. acknowledges support from Mineco
(No. FIS2013-41144-P) and
AGAUR (Grant No. 2014SGR878).
A. V. S. and P. T. acknowledge support from a bilateral German-Spanish
program of DAAD (project no. 57049473) via the
Bundesministerium f\"ur Bildung und Forschung (BMBF).
\scriptsize{
\bibliography{manus} 
\bibliographystyle{rsc} } 

\end{document}